\documentclass[a4paper,11pt]{article}

\usepackage{jheppub}
\usepackage[T1]{fontenc}
\usepackage{lmodern}
\usepackage{amsmath}
\usepackage{amsfonts}
\usepackage{amssymb}
\usepackage{mathrsfs}
\usepackage{physics}
\usepackage{graphicx}
\usepackage{caption}
\usepackage{subcaption}
\usepackage{enumitem}
\usepackage{framed,float}
\usepackage{booktabs}
\usepackage{array}
\usepackage{soul}
\usepackage{multirow}
\usepackage{comment}
\usepackage{braket}
\usepackage{todonotes}
\usepackage{youngtab}
\usepackage{tikz}
\usepackage{color}
\usepackage{extarrows}
\usetikzlibrary{calc,arrows,decorations.markings}
\usepackage{CJKutf8}
\usepackage{hyperref}

\graphicspath{{figures/}}

\def\e{\epsilon}

\def \be  {\begin{equation}}
\def \ee  {\end{equation}}
\def \ba {\begin{equation}\begin{aligned}}
\def \ea {\end{aligned}\end{equation}}
\def \bea  {\begin{eqnarray}}
\def \eea  {\end{eqnarray}}

\title{\boldmath Hot Holographic 2-flavor Quark Star}

\author[a]{Le-Feng Chen,}
\author[b,1]{Jing-Yi Wu,}
\author[a]{Hao Feng,}
\author[a]{Tian-Shun Chen,}
\author[a,c,d,1]{and Kilar Zhang\note{Corresponding author.}}

\affiliation[a]{Department of Physics and Institute for Quantum Science and Technology, Shanghai University, Shanghai 200444, China}
\affiliation[b]{School of Astronomy and Space Science, University of Chinese Academy of Sciences (UCAS), Beijing 100049, China }
\affiliation[c]{Shanghai Key Lab for Astrophysics, Shanghai 200234, China}
\affiliation[d]{Shanghai Key Laboratory of High Temperature Superconductors, Shanghai 200444, China}

\emailAdd{clf@shu.edu.cn}
\emailAdd{wujingyi222@mails.ucas.ac.cn}
\emailAdd{fenghaozi@shu.edu.cn}
\emailAdd{cts2003912@shu.edu.cn}
\emailAdd{kilar@shu.edu.cn}

\abstract{Applying the holographic 2-flavor Einstein--Maxwell-dilaton model, the parameters of which are fixed by lattice QCD, we extract the equations of state for hot quark--gluon plasma around the critical point at $T=182$ MeV, and have corresponding quark star cores constructed.  By further adding hadron shells,  the mass range of the whole stars spans from 2 to 17 solar masses, with the maximum compactness around {0.22}. This result allows them to be black hole mimickers and candidates for gap events. The I--Love--Q--C relations are also analyzed, which show consistency with the neutron star cases when the discontinuity at the quark--hadron interface is not large.  Furthermore, we illustrate the full parameter maps of the energy density and pressure as functions of the temperature and chemical potential and discuss the constant thermal conductivity case supposing a heat source inside.}

\begin{document} 
\maketitle
\flushbottom

%


\section{Introduction}\label{sec:1}
How to obtain bare quarks has been a long challenge,  since the extreme conditions are almost impossible to realize on the earth. A~possible solution may lie in compact stars, like in the cores of neutron stars (NSs).

Gravitational Waves (GWs) offered us a brand new way to detect NSs in the summer of 2017~\cite{LIGOScientific:2017vwq,LIGOScientific:2018hze,LIGOScientific:2020aai}, deciphering the information of masses and tidal deformations~\cite{Hinderer:2007mb}. Together with a traditional electromagnetic (EM) observation apparatus like NICER~\cite{Miller:2021qha}, we can narrow the windows and place constraints on the NS equation of state (EoS), excluding many EoS candidates. However,  observations alone are far from enough to break the degeneracy, and~theoretical derivations are~needed.

Unfortunately, theoretically obtaining the exact EoS for NSs is still an open question,  not to mention quark stars (QSs), 
require highly non-perturbative quantum chromodynamics (QCDs) beyond~the current analysis and calculation power.  At~this stage, holographic methods~\cite{witten1998anti,Polchinski:2000uf} provide hopeful approaches to extract the EoS. Two major holographic QCD models originating from string theory, one from brane constructions and~the other called Einstein--Maxwell-dilaton (EMD) \cite{Gibbons:1987ps} model, could both lead to competitive EoS~candidates.

For brane constructions, as~a top-down model, applying D4/D8 brane configurations, namely Witten--Sakai--Sugimoto (WSS) model~\cite{witten1998anti, Sakai:2004cn, Sakai:2005yt}, one can extract NS EoS. On~the other hand, D3/D7 configurations can lead to QS~EoS.

For QS itself, there are also two scenarios. Low or zero temperature and high density lead to quark matter, but~most of the current models (including D3/D7 model) \cite{Hoyos:2016zke,Hoyos:2016cob,Annala:2017tqz,BitaghsirFadafan:2019ofb,Zhang:2022uin} do not favor the existence of quark cores, as~the stars constructed are unstable. As~some exceptions, studies~\cite{BitaghsirFadafan:2020otb,Annala:2019puf,Han:2022rug,Fan:2023spm,Kurkela:2009gj} have found that stable NS can form with cold quark cores. High temperature and relatively lower density result in quark--gluon plasma, which may form a new type of star~\cite{Chen:2025caq} rather than~NS. 

{In} \cite{Rougemont:2023gfz}, {the holographic EMD model used to describe strongly coupled quark--gluon plasma is reviewed. The~EMD model can not only predict transport coefficients, but~also predict the existence of critical endpoint (CEP) in the QCD phase diagram. We adopt a model} \cite{Cai:2022omk} {introduced in this review because~its thermodynamic quantities can be obtained through holographic renormalization and thermodynamic relations, thereby extracting the EoS.} {Ref.~} 
\cite{Cai:2022omk} {is based on the DeWolfe--Gubser--Rosen (DGR)} \cite{DeWolfe:2010he} {model, and~improvements to this model enable {2+1} 
 flavor QCD matter at $\mu_B=0$ to quantitatively match the latest lattice data} \cite{HotQCD:2014kol,Borsanyi:2021sxv}, {thereby determining the precise coordinates of the critical endpoint at $(\mu_{CEP}, T_{CEP}) = (555~\rm{MeV}, 105~\rm{MeV})$ and characterizing the first-order transition line. The~2-flavor model} \cite{Zhao:2023gur} {adopts the coupling function form obtained from the 2+1-flavor case, using lattice data} \cite{Burger:2014xga} {from simulations performed at bare quark masses corresponding to pion mass $m_\pi\sim 360 \rm{MeV}$ and $N_t=12$, with~$N_f=2$ degenerate quark flavor. They set the pseudo-critical temperature from lattice simulations} \cite{Datta:2016ukp} {to $T_c(\mu_B=0) = 205~\rm{MeV}$ to match the lattice data, which falls within the deconfinement range of 219 ± 3 ± 14~MeV obtained from} \cite{Burger:2014xga}. Moreover, the~paper~\cite{Zhao:2023gur} calculates the relationship between baryon number density and temperature for different $\mu_B/T$ ratios. The~results show that the holographic predictions match well with the lattice results at low chemical potential~\cite{Datta:2016ukp}, which strongly supports the hQCD model we employed. We considered $2+1$-flavor case in~\cite{Chen:2025caq}, and~now we discuss the $2$-flavor~case. 

There are two major differences from the last work. First is, of course, the change in flavors. Constraining to only up and down quarks is, in some sense, more realistic. 
It is noteworthy that the change in quark flavor number has a very pronounced effect on the CEP position. Applying the same EMD model, in the case of 2+1-flavor, CEP is located at $(\mu_{CEP}, T_{CEP}) = (219~\rm{MeV}, 182~\rm{MeV})$ \cite{Cai:2022omk},  while for 2-flavor, $(\mu_{CEP}, T_{CEP}) = (555~\rm{MeV}, 105~\rm{MeV})$ \cite{Zhao:2023gur}. {The determining procedure is briefly as follows:
Through calculations of the temperature dependence of the Polyakov loop and free energy density, the~EMD model finds that when $\mu$ approaches the critical value (219~MeV), the~susceptibility of the Polyakov loop becomes infinite. Beyond~this critical value, both the Polyakov loop and free energy density exhibit multi-valued behavior, indicating the occurrence of a first-order phase transition, which also determines the position of the CEP. In~practice, the~form of the potentials} \eqref{eq z} {are the same for both 2+1-flavor and 2-flavor cases, but~the coefficients are set differently according to the corresponding lattice data (as illustrated in Table~1 of} \cite{Zhao:2023gur}), {thus leading to different CEP. 
The reason why the 2+1-flavor model is sometimes favored is that the~state with a strange quark is considered to be the ground state of baryonic matter at zero temperature in the literature }\cite{Witten:1984rs}. {However, in~2018, there was also a claim that the 2-flavor case can be more stable if~the baryon number is more than 300 (although, this is beyond the periodic table)} \cite{Holdom:2017gdc}.  {Those two conclusions are for cold quark matter, while for hot quark--gluon plasma, constraining to only up and down quarks is more natural, without~the need of introducing strange quarks.
}

Secondly, in~contrast to previous studies that computed the interior structures of finite-temperature QS and NS models using either the isothermal assumption or other simplified relations~\cite{Blaschke:1998hy,Maiti:2024dbl,Kettner:1994zs}, we now show the whole parameter spaces for $\e$ and $p$
as functions of T and $\mu$ and,  in~addition~discuss the special curves under constant thermal conductivity.

We end up with hot and massive QS, {the masses of which range from 2 to 17 $\text{M}_{\odot}$, which are much larger than NS, though~lower compared to their 2+1-flavor counterparts with 5 to 30~$\text{M}_{\odot}$} in~\cite{Chen:2025caq}. This property enables them to mimic black holes (BHs) rather than~NS. 

For the formation and lifetime of such high temperature stars, similar questions arise like the cases in~\cite{Chen:2025caq}. In~both the early universe and core-collapse supernovae, extreme conditions may generate temperatures sufficient for quark--gluon plasma formation. While supernovae can reach as high as 30--50 MeV~\cite{Sekiguchi:2011mc, Drago:2015dea} in the literature, we suggest that certain extreme explosions could attain temperatures several times higher, to more than 182~MeV, potentially creating hot QS. While hot QS may predict rapid collapse into NS and subsequently BH due to intense thermal radiation, the~models in this paper are dealing with a stratified structure featuring a quark-phase core enveloped by neutron-rich matter. Crucially, strong radiation reflection at the phase transition boundary may significantly reduce energy leakage from the quark core, thereby extending the stellar lifespan. Detailed investigation of these interface dynamics and the observational implications will be addressed in future~work.

In order to analyze the ability of detections, we calculate the universal relations across the moment of inertia (I), the~tidal deformability (Love), and~the quadrupole moment (Q) and the compactness (C), which is also called the I--Love--Q--C relations. The~relations have been found to be independent on most models, whether the EoS are continuous~\cite{Yagi:2013bca, PhysRevD.88.023009,Silva2017ILoveQTT,PhysRevD.103.023022,Li:2023owg,Atta:2024ckt,Kumar:2023ojk,Pani:2015tga} or have small discontinuity~\cite{Martinon_2014,Lau_2019,Raduta_2020,PhysRevD.105.104018,ROY2025103108}. As~BH mimickers, the~stars are distinguishable by comparing the masses with the detection results of NS. Furthermore, the~stars can be differentiated from BH using characteristic wave forms and,~further, the non-zero tidal love number (TLN) in contrast to the zero TLN of BH in general relativity. The~deviations of the I--Love--Q--C relations for different models are also effective in observations. Moreover, the~thermal radiation emitted by such high-temperature stars provides traditional methods for their~detection.

This paper is organized as follows. Section~\ref{sec:1} is the introduction, and~Section~\ref{sec:2} summarizes the way to extract EoS from the holographic 2-flavor EMD model. Section~\ref{sec:iloveq} shows the QS constructed and analyzes their I--Love--Q--C relations. We discuss more on the full parameter space of $\e$ and $p$ and~consider a constant thermal conductivity case in Section~\ref{sec:5}. We provide conclusions in Section~\ref{sec:6}.

\section{Holographic 2-Flavor QCD~Model}\label{sec:2}
Research was conducted within the EMD theoretical framework in five-dimensional space. The~action of the system is represented as:
\begin{equation}
    \begin{split}
        S=&\frac{1}{2\kappa_N ^2}\int d^5x\sqrt{-g} \left[R-\frac{1}{2}\bigtriangledown_\mu\phi\bigtriangledown^\mu\phi-\frac{Z(\phi)}{4}F_{\mu\nu} F^{\mu\nu}-U(\phi)\right], \label{eq 1}
    \end{split}
\end{equation}
where $\kappa_N$ represents the effective Newton constant, $\phi$  is a scalar field called dilaton, which breaks conformal symmetry, $F_{\mu\nu}$ is the field strength tensor of the U(1) gauge field, and~g denotes the determinant of the five-dimensional spacetime metric tensor, R is the Ricci scalar curvature. The~scalar potential $U(\phi)$ and the gauge coupling function $Z(\phi)$ are chosen to reproduce key features of 2-flavor lattice QCD at zero chemical potential. Based on the work of~\cite{Zhao:2023gur}, we adopt the {following forms:} 
\begin{equation}
\begin{split}
    U(\phi)&=-12\,{\rm cosh}[c_1 \phi]+(6c_1^2-\frac{3}{2})\phi^2+c_2 \phi^6, \\
    Z(\phi)&=\frac{{\rm sech}[c_4 \phi^3]}{1+c_3}+\frac{c_3}{1+c_3}e^{-c_5\phi}.\label{eq z}
    \end{split}
\end{equation}

{As established} 
 by lattice data in~\cite{Zhao:2023gur}, the~parameters $\kappa_N$ and $c_i$ are determined as the following values: $\kappa_N^2=2\pi(3.72)$, $c_1=0.7100$, $c_2=0.0002$, $c_3=0.530$, $c_4=0.085$, and~$c_5=30$.

The metric form of a five-dimensional AdS black hole with scalar hair is as follows
\begin{equation}
    ds^2=-f(r)e^{-\eta(r)}dt^2+\frac{dr^2}{f(r)}+r^2(dx^2+dy^2+dz^2),
\end{equation}
and
\begin{equation}
    \phi=\phi(r), \quad A_t=A_t(r).
\end{equation}
where $f(r)$ and $\eta(r)$ are all functions of r, and~$A_t$ is the t-component of the gauge~field.

The temperature $T$ and entropy density $s$ can both be obtained from the horizon
\begin{equation}
   T=\frac{1}{4\pi}f'(r_h)e^{-\eta(r_h)/2}, s=\frac{2\pi}{\kappa_N ^2}r^3_h \label{eq1},
\end{equation}
where $r_h$ is the horizon~radius.

Through the calculation method in~\cite{Cai:2022omk}, the~specific form of energy density $\e$ and pressure $p$ can be obtained
\begin{equation}
\begin{split}
    \epsilon&=\frac{1}{2\kappa_N^2}(-3f_v+\phi_s \phi_v+\frac{1+48b}{48}\phi_s^4),\\
    p&=\frac{1}{2\kappa_N^2}(-f_v+\phi_s \phi_v+\frac{3-48b-8c_1^4}{48}\phi_s^4),\label{eq 2}
\end{split}
\end{equation}
where $f_v,~\phi_s,~\phi_v$ are the parameters in the asymptotic expansion of the metric function on the AdS boundary, and~$b=-0.25707$ is the result from holographic renormalization. The~$\phi_s$ originates from the asymptotic expansion of scalar fields, and~their non-zero values break the conformal symmetry. Additionally, the~value of $\phi_s=1.227~\text{GeV}$ \cite{Zhao:2023gur} can be fixed through lattice~QCD.

To obtain the mass and the radius of the star through solving Tolman--Oppenheimer--Volkoff (TOV) equations with the EoS, we need to convert the usual QCD units [$\rm{MeV}^4$] into the so-called astronomical units $r_{\odot}$, $\epsilon_{\odot}$, $p_{\odot}$, with~$r_{\odot}=G_N M_{\odot} / c^2, \epsilon_{\odot}=M_{\odot} /r_{\odot}^3$, and~$p_{\odot}=c^2 \epsilon_{\odot}$,
\begin{align} 
    \epsilon_{\odot}&=\textit{p}_{\odot}=6.60271\times10^{14}\,\rm{MeV^4}=3.14644\, \times10^5~\rm{MeV\cdot fm^{-3}},
\end{align}
where we have taken $c=G_N=1$.

\begin{figure}[H]
\includegraphics[width=13.5 cm]{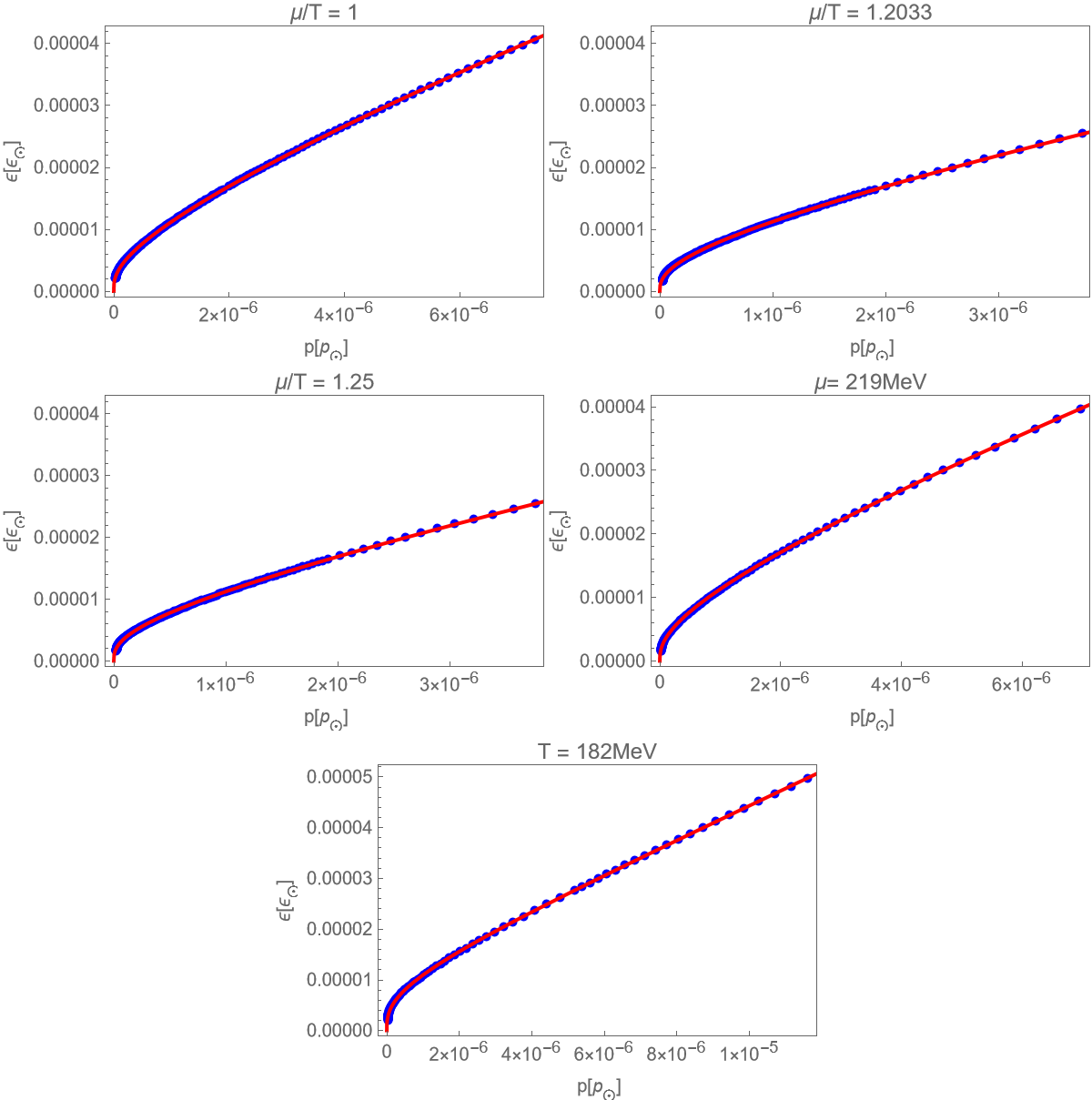}
\caption{{Energy--pressure} 
 data and fitted curves under five different ${T,\mu}$ conditions. Blue points represent numerical data, and~red lines are fitted curves. Here, our data does not reach zero because this is high-temperature quark matter, which still maintains a certain pressure at the phase transition~point.}
\label{fig:fitting}
\end{figure}

In~\cite{Zhao:2023gur}, they obtained the critical end point (CEP) of 2-flavor QCD $(\mu=219~\text{MeV}$, $\mbox{\emph{T}~=~182~\text{MeV}})$. For~simplified analysis, we selected five sets of data points around CEP, which are $\mu/T=1$, $\mu/T=1.2033$, $\mu/T=1.25$, $T=182~\rm{MeV}$ and $\mu=219~\rm{MeV}$, and~performed fittings on these data, with~the fitting results and data shown in Figure~\ref{fig:fitting}. The~real EoS will be contained within this range, though~complicated to obtain~directly.

We must emphasize that the~blue data points in Figure~\ref{fig:fitting} do not reach zero because we select the high-temperature quark--gluon plasma, in which case neither pressure nor temperature is zero.

The fitting results corresponding to different physical conditions are listed below:
$p_1, \epsilon_1$ for $\mu/T=1$,
$p_2, \epsilon_2$ for $\mu/T=1.2033$,
$p_3, \epsilon_3$ for $\mu/T=1.25$,
$p_4, \epsilon_4$ for $T=182\ \mathrm{MeV}$,
$p_5, \epsilon_5$ for $\mu=219\ \mathrm{MeV}$.
\begin{equation}
    \begin{split}
        \epsilon_1&=0.000501863 \, p_1^{0.315468} + 1.15292 \,p_1^{0.89651}, \\
        p_1&\in [3.0918\times 10^{-8}, \;7.3047\times 10^{-6}], \\ 
        \epsilon_1&\in[2.1682\times 10^{-6},\; 4.066\times 10^{-5}], 
    \end{split}
    \label{eq:eos1} 
\end{equation}
\begin{equation}
    \begin{split}
        \epsilon_2&=0.00139264 \,p_2^{0.375067}+2.6989 \,p_2^{0.982421},\\ 
        p_2&\in [2.627\times 10^{-8}, \;7.3045\times 10^{-6}], \\ 
        \epsilon_2&\in[1.7023\times 10^{-6},\; 4.0658\times 10^{-5}],
    \end{split}
    \label{eq:eos2}
\end{equation}
\begin{equation}
    \begin{split}
        \epsilon_3&=0.000928211 \,p_3^{0.349754} +2.15744 \,p_3^{0.958067},\\ 
        p_3&\in [2.3732\times 10^{-8}, \;7.35366\times 10^{-6}],\\ 
        \epsilon_3&\in[1.6962\times 10^{-6}, \;4.08\times 10^{-5}], 
    \end{split}
    \label{eq:eos3}
\end{equation}
\begin{equation}
    \begin{split}
        \epsilon_4&=0.00117483 \,p_4^{0.350591} +11.3855 \,p_4^{1.13679},\\ 
        p_4&\in [2.61368\times10^{-8},\; 1.1615\times10^{-5}], \\ 
        \epsilon_4&\in[2.1067\times10^{-6}, \;4.9673\times 10^{-5}], 
    \end{split}
    \label{eq:eos4}
\end{equation}
\begin{equation}
    \begin{split}
        \epsilon_5&=0.001427 \,p_5^{0.378816} +2.09627 \,p_5^{0.958741}, \\ 
        p_5&\in [2.60394\times10^{-8}, \;6.97167\times10^{-6}], \\ 
        \epsilon_5&\in[1.5845\times10^{-6}, \;3.967\times 10^{-5}]. 
    \end{split}
    \label{eq:eos5}
\end{equation}

The permissible fitting ranges for pressure ($p$) and energy density ($\epsilon$) in astrophysical units are constrained as follows: the minimum values correspond to the phase transition threshold, whereas the maximum values are determined by stellar stability~requirements.

\section{Hot Quark Stars and Their I--Love--Q--C~Relations}\label{sec:iloveq}
Theoretically inferred models play a crucial role in supporting astrophysical observations by leveraging universal relations among key macroscopic properties of compact stars—namely, the I--Love--Q trio~\cite{Yagi:2013bca,PhysRevD.88.023009}. These nearly EoS-independent relations enable the extraction of fundamental stellar parameters from gravitational wave and pulsar timing observations, providing a powerful tool for probing the internal structure and composition of compact objects, including NS and~QS.

\begin{figure}
    \centering
    \includegraphics[width=\linewidth]{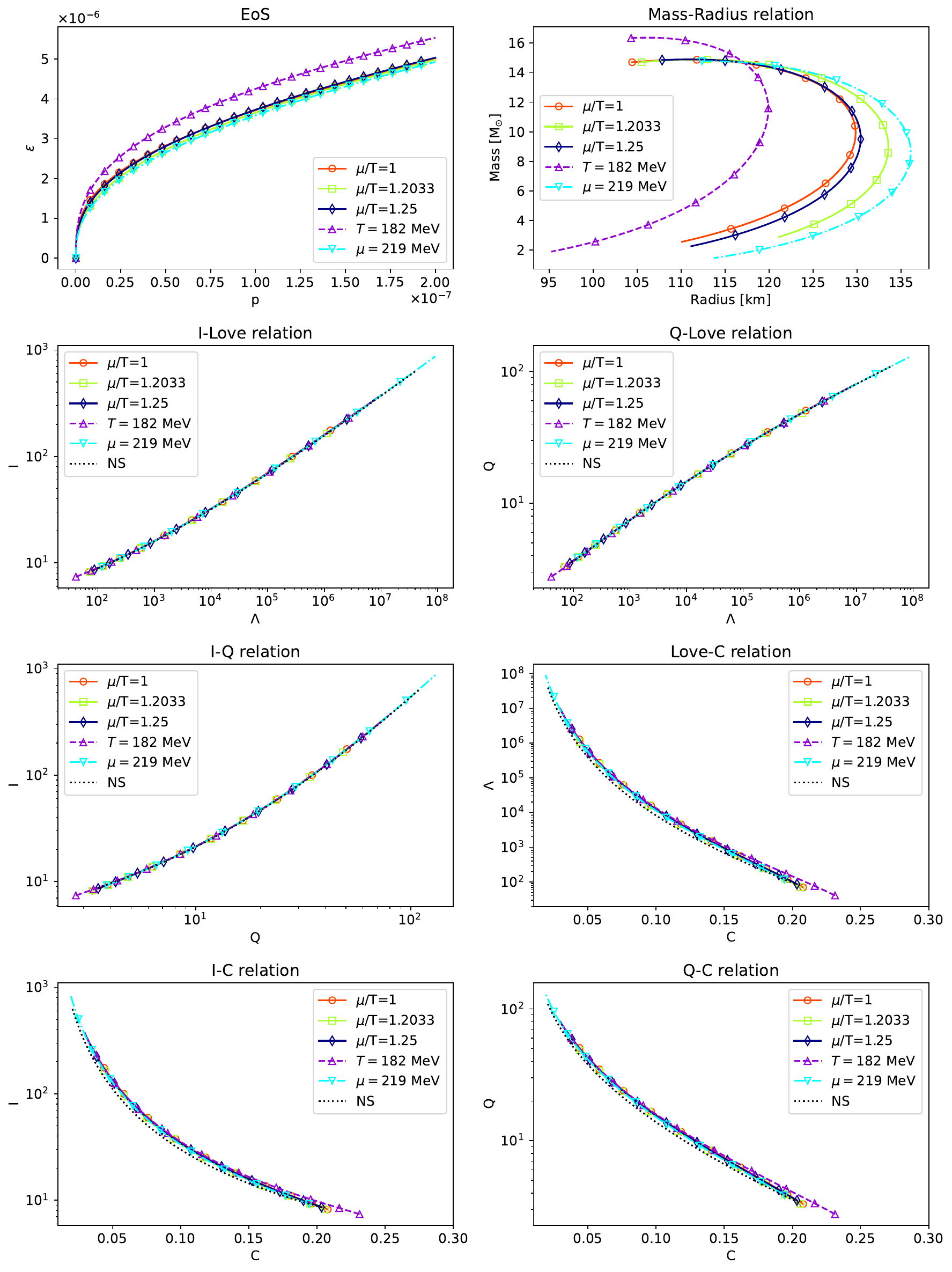}
    \caption{{The} 
 EoS, M--R relation, and~I--Love--Q--C relations for the ``simple'' QS are presented using various models described in Equations~(\ref{eq:eos1})--(\ref{eq:eos5}). For~comparison, the~panels showing the I--Love--Q--C relations of a polytropic NS, given by $\epsilon_n = 0.09p_n^{0.5}$,
    is plotted with a black dotted line.}
    \label{iloveqc}
\end{figure}

In Figure~\ref{iloveqc}, we plot the mass--radius (M-R) curve and the I--Love--Q--C relations of the models in Equations~(\ref{eq:eos1})--(\ref{eq:eos5}) with the calculation procedures in~\cite{PhysRevD.88.023009}. We plot the $\epsilon$ and $p$ with TOV units $\epsilon_{\odot}$ and $p_{\odot}$ while the I--Love--Q--C are dimensionless. We should note that, as shown in Figure~\ref{fig:fitting}, the~pressure of the quark phase is nonzero, but~for the simplest analysis, we extend the fitting of the models to $p=0$ as the surface condition required, so that the hadron crust shares the same EoS curve (with a smooth phase transition), and~we use the terminology ``simple'' to represent the models. For~comparison, we 
plot the I--Love--Q--C relations for a single polytropic NS model
\begin{equation}
    \epsilon_{n} = \kappa_{n} \,p^{\gamma_{n}},
    \label{eq:nseos}
\end{equation}
with the black dotted curve. Here, $\gamma_{n}$ is set to 0.5 as a typical reference and the corresponding $\kappa_{n}=0.09$, which satisfies the constraint that the maximum mass of NS is about $2.2~\text{M}_{\odot}$, which are also the parameters used in~\cite{Wu:2023aaz}. From~Figure~\ref{iloveqc}, we observe that the maximum masses of the quark models can reach up to $17~\text{M}_{\odot}$. In~sight of the panels of the I--Love--Q trio, we find the models still obey the universal relations, compared with the NS model. But~for the curves about the compactness, the~universality is broken. We regard the deviations as plausible because there exist differences between the unrealistic NS EoS and the realistic ones~\cite{PhysRevD.88.023009}. 

More realistically than the ``simple'' EoS cases, we introduce the ``combined'' cases:  considering that the quark--gluon plasma cools down as it approaches the surface and eventually undergoes a phase transition, we connect the quark cores with the hadron (probably, neutron) crusts at their respective minimum pressures. The~hadron model we use is Equation~(\ref{eq:nseos}) with a fixed $\gamma_n=0.5$. At~the transition point, we set
\begin{equation}
    \epsilon_{n}(p_t) = m \,\epsilon_{q}(p_t),
    \label{phasetran}
\end{equation}
to illustrate the discontinuities. Here, $\epsilon_q$ is referred to as the energy density of the quark cores, and $p_t$ represents the transition point. The~parameter $\kappa_{n}$ depends on the choice of $m$. 

\begin{figure}
    \centering
    \includegraphics[width=\linewidth]{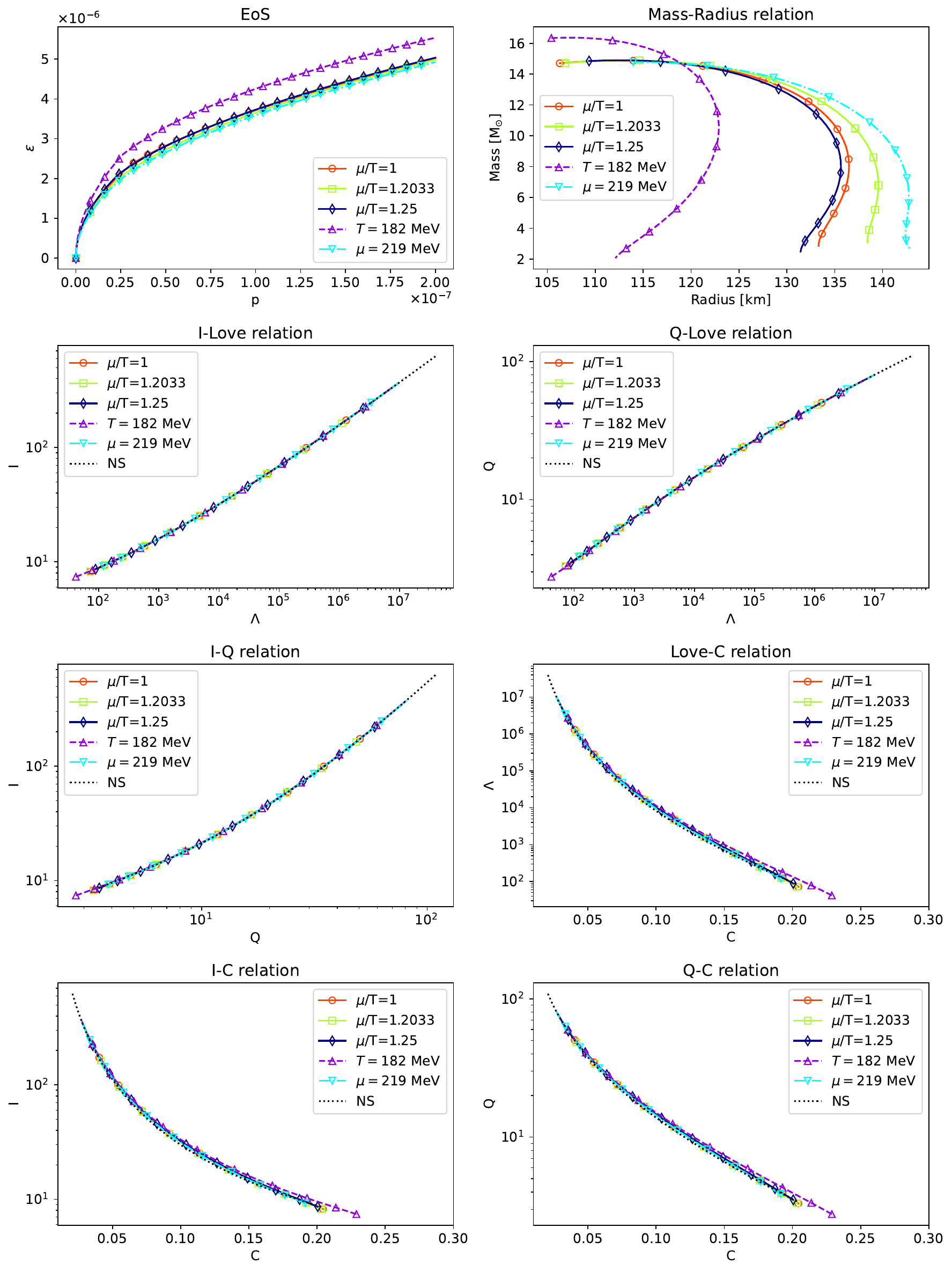}
    \caption{{The} 
 EoS, M--R relation, and~I--Love--Q--C relations for the ``combined'' QS consisting of quark cores and hadron shells with $m=1$ in Equation~(\ref{phasetran}). For~comparison, the I--Love--Q--C relations of a polytropic NS, given by $\epsilon_n = 0.09p_n^{0.5}$,
    are plotted with a black dotted line.}
    \label{iloveqc1}
\end{figure}

\begin{figure}
    \centering
    \includegraphics[width=\linewidth]{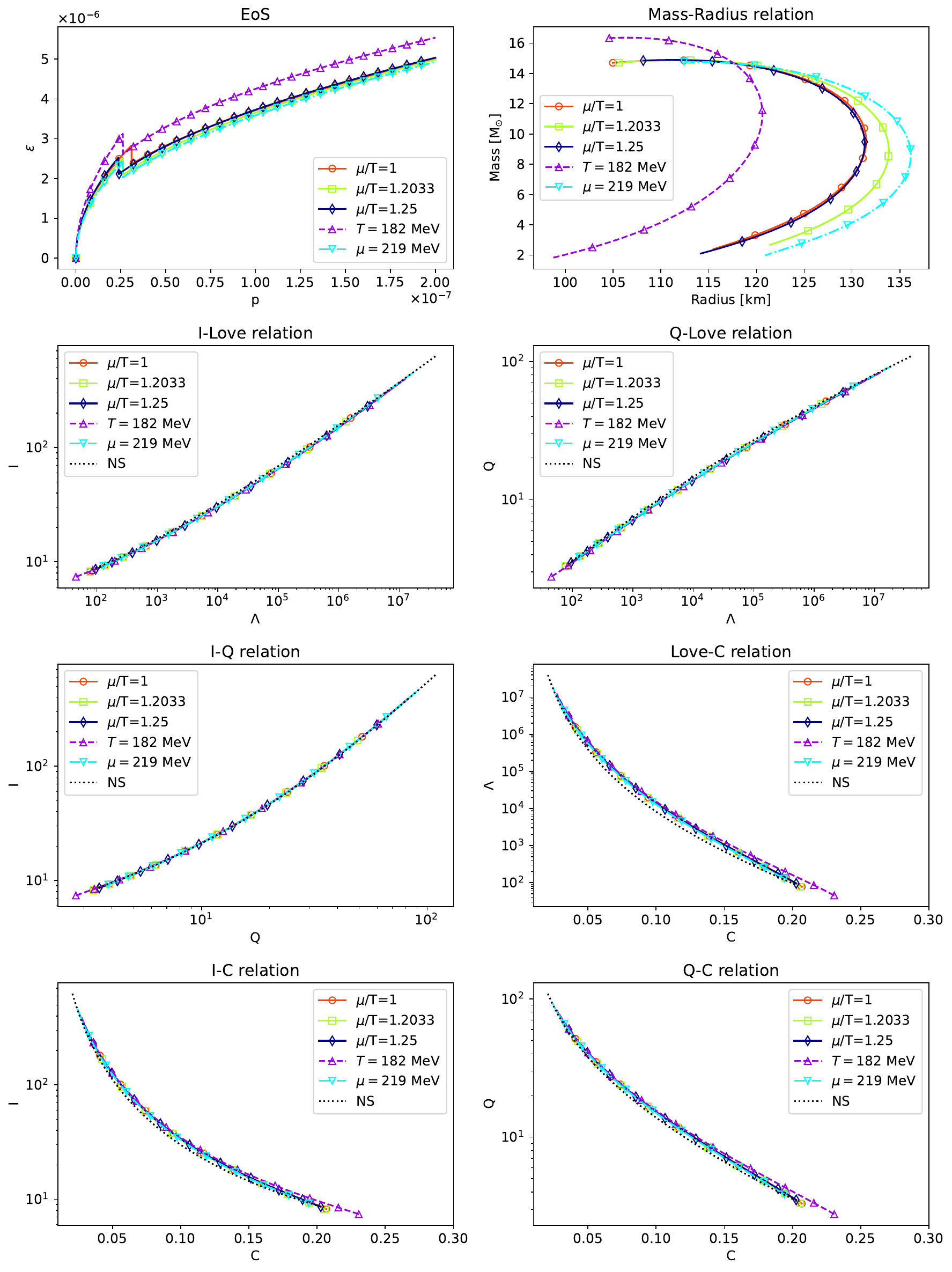}
    \caption{{The} 
 EoS, M--R relation, and~I--Love--Q--C relations for the ``combined'' QS consisting of quark cores and hadron shells with $m=1.2$ in Equation~(\ref{phasetran}). For~comparison, the I--Love--Q--C relations of a polytropic NS, given by $\epsilon_n = 0.09p_n^{0.5}$,
    are plotted with a black dotted line.}
    \label{iloveqc12}
\end{figure}

\begin{figure}
    \centering
    \includegraphics[width=\linewidth]{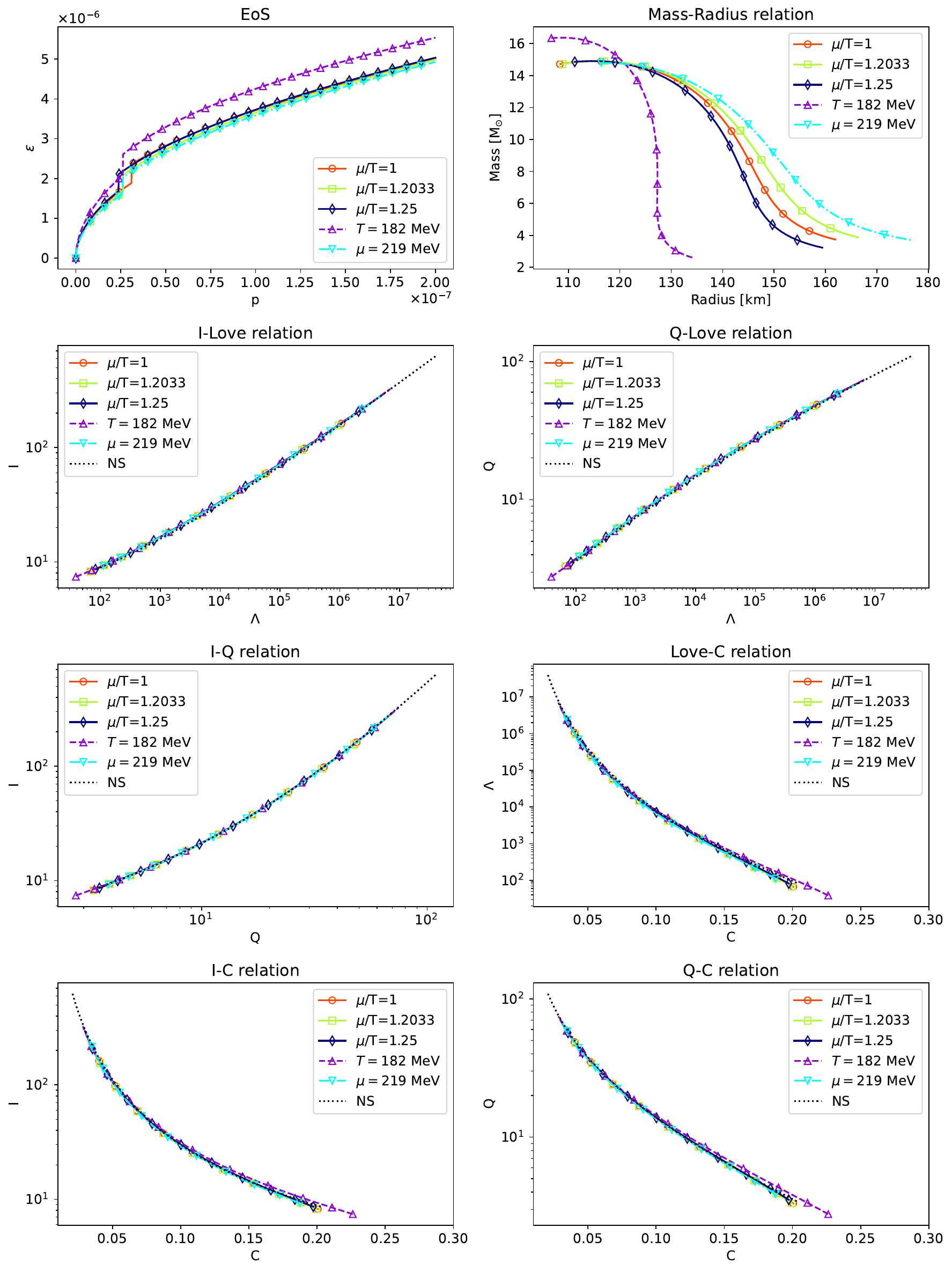}
    \caption{{The} 
 EoS, M--R relation, and~I--Love--Q--C relations for the ``combined'' QS consisting of quark cores and hadron shells with $m=0.8$ in Equation~(\ref{phasetran}). For~comparison, the the I--Love--Q--C relations of a polytropic NS, given by $\epsilon_n = 0.09p_n^{0.5}$,
    are plotted with a black dotted line.}
    \label{iloveqc08}
\end{figure}

\begin{figure}
    \centering
    \includegraphics[width=\linewidth]{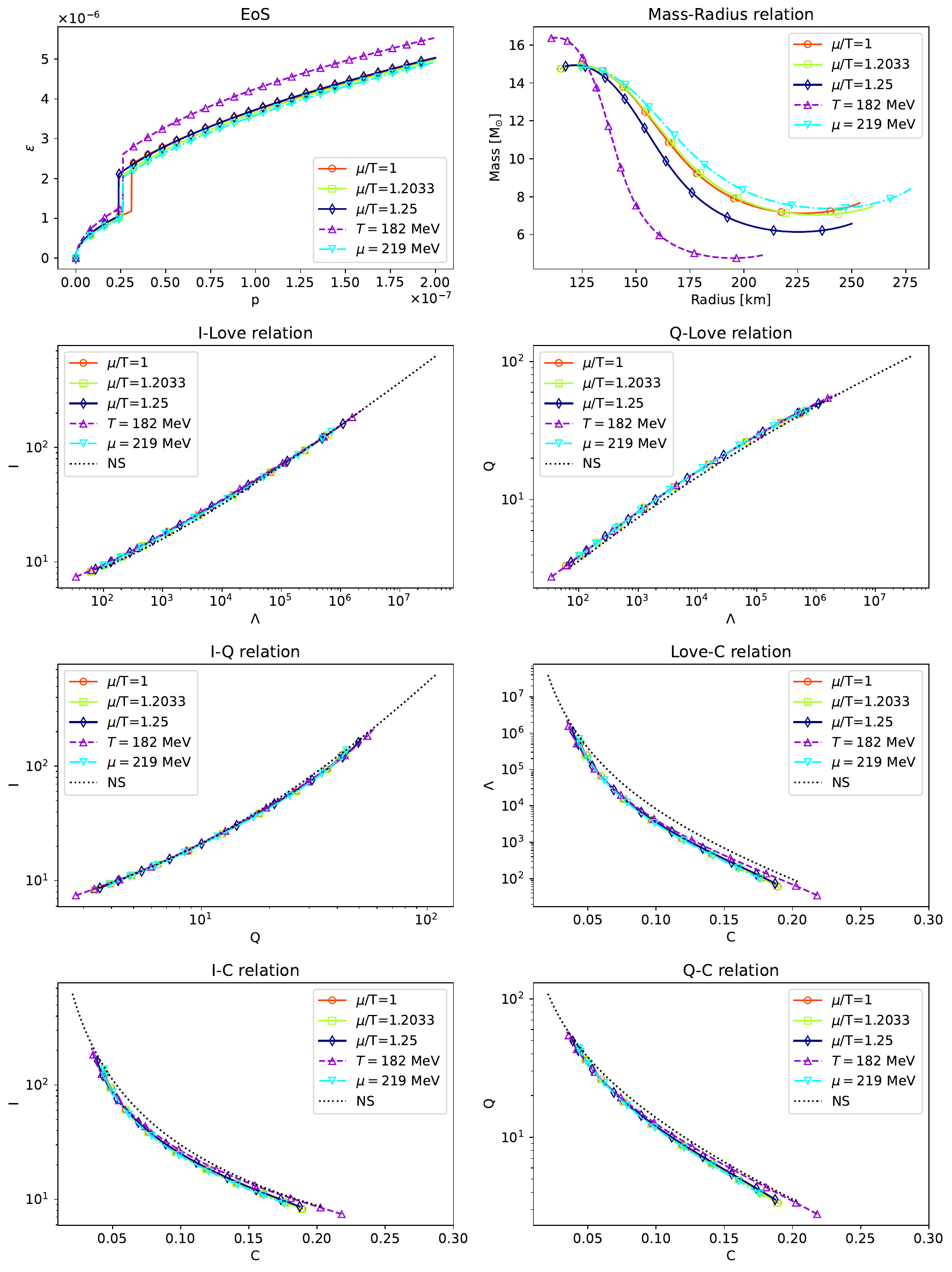}
    \caption{{The} 
 EoS, M--R relation, and~I--Love--Q--C relations for the ``combined'' QS consisting of quark cores and hadron shells with $m=0.5$ in Equation~(\ref{phasetran}). For~comparison,  the I--Love--Q--C relations of a polytropic NS, given by $\epsilon_n = 0.09p_n^{0.5}$,
    are plotted with a black dotted line.}
    \label{iloveqc05}
\end{figure}

We plot the EoS, M--R relations and I--Love--Q--C relations for ``combined'' QS with $m=1$, $1.2$, $0.8$ and $0.5$ in Figures~\ref{iloveqc1}--\ref{iloveqc05} through the procedures for the two-layer models in~\cite{Zhang:2020pfh}.
In the panels for M--R relations, we show that the ``combined'' QS have more unstable components (i.e., the higher radii parts of the stars) for the larger phase transitions, while the stability judging methods can be seen in~\cite{1966ApJ...145..514M}. We should mention that the selection of the ranges of the central pressure $p_0$, which are the initial conditions when solving TOV equations to derive the mass and radius, are the same across all ``combined'' QS for the respective quark EoS. Through the trend of instability with the phase transition, we can foresee that the stars will not exist in extreme circumstances.
From the lower six panels, we observe that the I--Love--Q relations are valid for all the ``combined'' cases. The~I--C, Love--C and Q--C curves obey the universality except for $m=0.5$, which may be induced by the larger phase transition. Compared with Figure~\ref{iloveqc}, the~relations for the ``combined'' cases conform better. The~reason is the influence from the hadron shells, which is more dominant for the universality than the cores~\cite{PhysRevD.88.023009,Wu:2023aaz}.

Compared with the 2+1-flavor model~\cite{Chen:2025caq}, we obtain nearly half of the masses of the stars from the present models, correspondingly with  half of the radii. Furthermore, the~universal relations of the 2-flavor models seem to deviate from the referred NS model less than our previous work. The~result reduces the mass gaps between NS and the QS in our previous work. 
We think the observations on the 2-flavor QS are able to supply the formation theories of the compact~stars.

\unskip

The ranges of the masses of the ``simple'' QS and ``combined'' cases enable these compact objects to emulate the observable properties of stellar-mass BH. The~present study demonstrates that the I--Love--Q--C universal relations remain valid across a broad class of ``combined'' configurations, thereby reinforcing their applicability beyond 2-flavor quark matter systems. From~an observational perspective, these universal relations may serve as a diagnostic tool to discriminate between BH and QS, particularly due to the vanishing TLN of BH in general relativity. Additionally, the~observed deviations in compactness-related relations may offer potential insights into holographic quark matter models. Importantly, observations of ``combined'' QS could constrain the parameters governing the phase transition between deconfined quark matter and hadronic matter, thereby offering a pathway to probe the microphysics of dense matter in astrophysical~environments.


\section{More Discussions on the \boldmath{$\epsilon$} and \boldmath{$p$} Parameter~Space}\label{sec:5}

In this section, we illustrate the whole $\epsilon$ and $p$ parameter space, and~also consider a more complicated case study by replacing the previous simple conditions on $\mu$ and $T$ (as discussed in Section~\ref{sec:2}) with a constant thermal conductivity~constraint.

\begin{figure}[H]
    \includegraphics[width=13.5 cm]{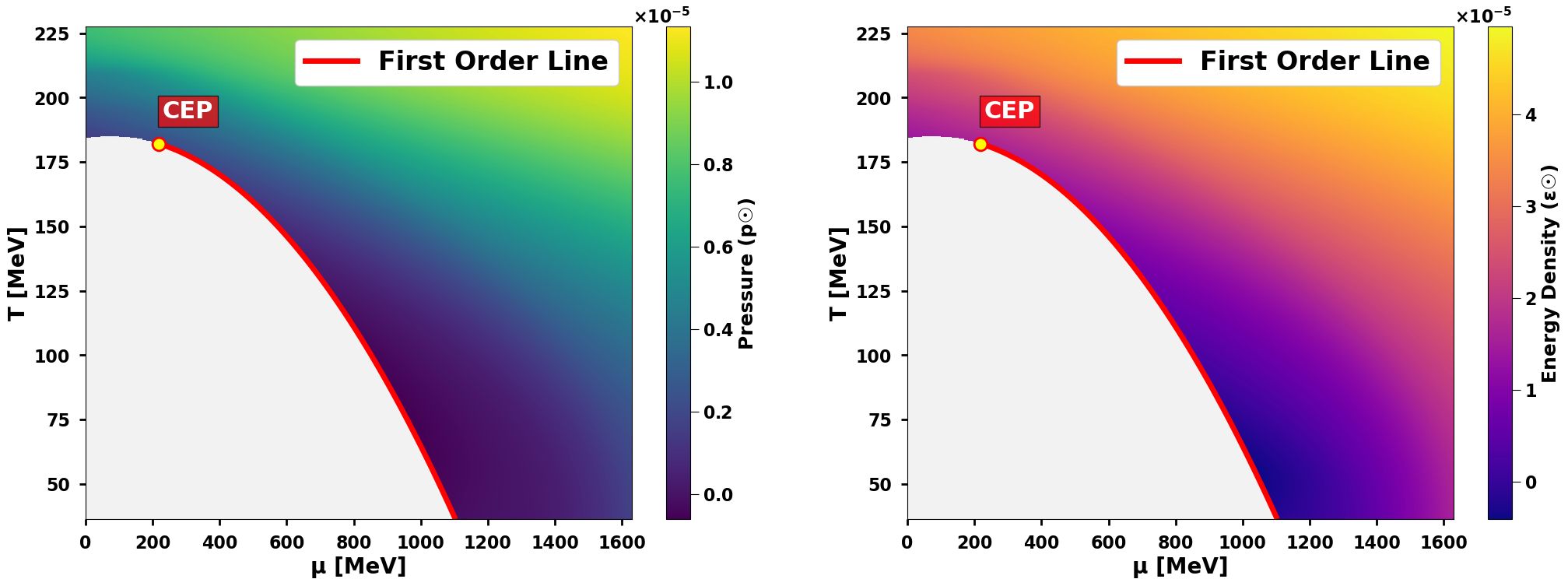}
    \caption{{The} 
 heatmap presents all the $T$, $\mu$ dependence for $p$ and $\epsilon$ of the Holographic 2-flavor QCD Model (notice that here for $p$ and $\epsilon$ we are using astronomical units $p_{\odot}$ and $\epsilon_{\odot}$ for later convenience, rather than QCD units): $p = p(\mu, T)$ and $\epsilon = \epsilon(\mu, T)$, where the shading of the background color represents the magnitudes of $p$ and $\epsilon$. The~CEP and first-order line are taken from reference~\cite{Zhao:2023gur}.
    }
    \label{EoS}
\end{figure}

First, we numerically determine the relationships $\epsilon(\mu,T)$ and $p(\mu,T)$ using a holographic 2-flavor QCD model, presenting these correlations as heatmaps in Figure~\ref{EoS}. 

This visualization displays all the EoS information for the quark phase of a holographic 2-flavor QCD model, expressing pressure $p$ and energy density $\epsilon$ as functions of chemical potential $\mu$ and temperature $T$. The~color gradient quantitatively indicates the values of these thermodynamic quantities. Our investigation concentrates solely on the quark-dominated regime situated above the first-order transition boundary. In~the diagram, the~shaded gray area corresponds to the hadronic matter domain, with~both the CEP and first order line taken from reference~\cite{Zhao:2023gur}.

Typically, realistic conditions assume rapid attainment of thermal equilibrium, corresponding to the constant-temperature stellar scenario already addressed in the preceding discussion of $T=\text{constant}$.
We now turn to the next simplest non-trivial configuration---a stellar body with constant thermal conductivity containing a thermal reservoir in the center, which will be our focus in the following~analysis.

Although theoretically the thermal conductivity of each point in the $(\mu,T)$ parameter space could be derived from holographic principles, we defer the complete holographic analysis to future work. For~computational simplicity, we instead solve the TOV equations starting from static field equations under constant thermal conductivity conditions. The~temperature profile of QS is combined with the $\mu-$ and $T-$dependent expressions of energy density $(\epsilon)$ and pressure $(p)$ to obtain solutions, followed by a discussion of key~results.

The temperature distribution in stars with constant thermal conductivity satisfies the Laplace equation:
\begin{equation}
   \quad \nabla^2 T = 0\label{eq:laplace}.
\end{equation}
Under assumptions of spherical symmetry, the~solution takes the form $T = C/R$. However, this formulation does not yield a finite central temperature. To~resolve this divergence, we introduce a cutoff at radius $R = R_c$, establishing the temperature profile:
\begin{equation}
T(r) = \begin{cases} 
T_0 & r \leq R_c \\
\frac{T_0 R_c}{r} & r > R_c .
\end{cases}\label{eq:T-profile}
\end{equation}
Here, we suppose that the core region $(r \leq R_c)$ contains a constant-temperature heat source where the temperature field is not governed by the Laplace~equation.

Substituting $T(r)$ into $p(\mu,T)$ yields $\mu(p,r)$, which when combined with $\epsilon(T,\mu)$ produces $\epsilon(p,r)$. This relationship serves as the EoS for TOV equation~integration. 

\begin{figure}[H]
  \begin{subfigure}[t]{0.45\textwidth}
    \centering
    \includegraphics[width=\textwidth]{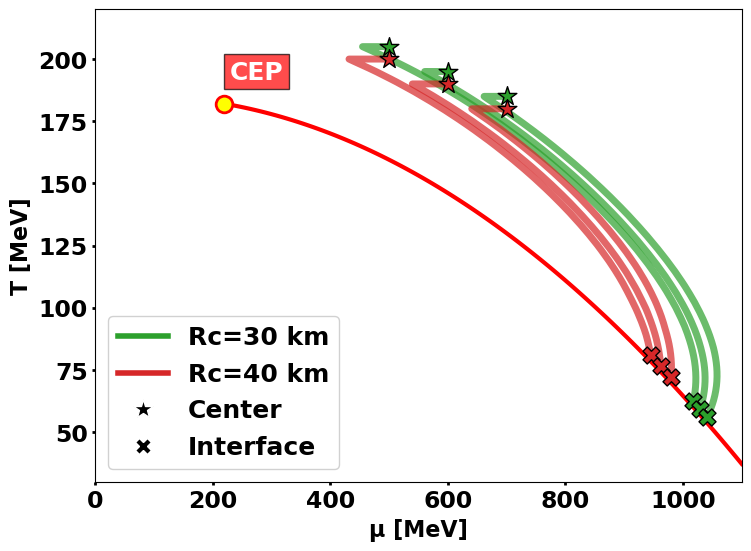}
    \caption{ \centering} 
    \label{fig:rc=30}
  \end{subfigure}
  \hfill
  \begin{subfigure}[t]{0.45\textwidth}
    \centering
    \includegraphics[width=\textwidth]{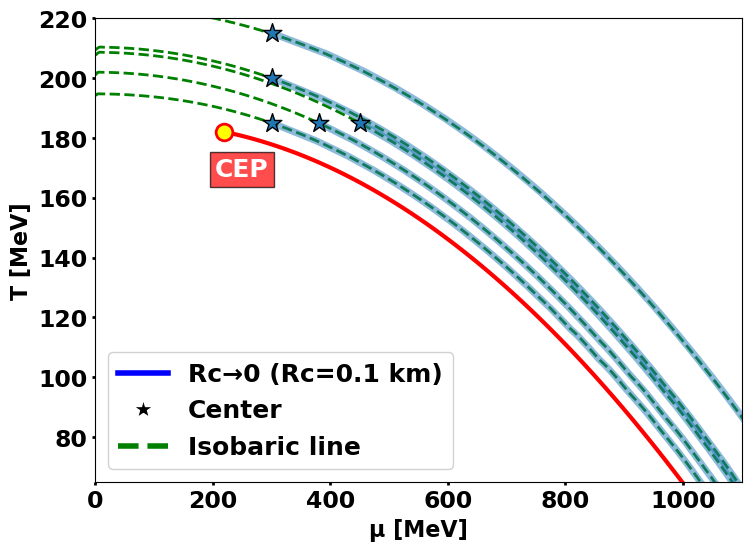}
    \caption{ \centering} 
    \label{fig:rc=0.1}
  \end{subfigure}

  \caption{  Paths of chemical potential and temperature ($\mu, T$) as functions of radius for different thermal core radius $R_c$:
    (\textbf{a}) Finite thermal core radius: As radius $r$ increases, the~path intersects the phase transition line and terminates at the stellar interface. Different central conditions ($\mu_0, T_0$) correspond to distinct trajectories, reflecting variations in the stellar EoS.
    (\textbf{b}) Infinitesimal core ($R_c \to 0$): The pressure remains approximately constant during radial expansion, resulting in an isobaric trajectory.}
  \label{fig:paths}
\end{figure}

Figure~\ref{fig:paths}a shows the evolutionary paths of $\mu(T)$ under TOV equation integration for different central initial conditions $(\mu_0, T_0)$ near the phase transition boundary with \mbox{$R_c = 30$ km} and $40$ km. 
The pentagrams denote the integration starting points (stellar centers), while the crosses mark the termination points at the phase transition~interface.

Notably, different initial conditions lead to distinct integration paths, which means that our approach differs fundamentally from classical solutions as the EoS becomes dependent on central temperature $T_0$. Consequently, each QS requires a unique EoS, rendering traditional NS analysis methods (e.g., M--R curve stability criteria) inapplicable~here.

We further investigate the $R_c\to 0$ limit to eliminate core effects in Figure~\ref{fig:paths}b. In~this regime, the~quark phase radius contracts proportionally with $R_c$, leading to constant pressure where $\mu(T)$ reduces to an isobaric curve. Remarkably, sufficiently compact stars exhibit pressure-dominated behavior regardless of thermal profile, rendering the constant thermal conductivity condition trivial in this~limit.

For physical QS, the~central temperature $T_0$ and chemical potential $\mu_0$ must satisfy constraints derived from matching to hadron phase exteriors. We adopt the same method as above by connecting to the external hadron shell EoS $\epsilon_{n} = \kappa_{n} \,p^{\gamma_{n}}$ (with $\gamma_{n}=0.5$) beyond the phase transition~boundary.

\begin{figure}[H]
    \includegraphics[width=13.5 cm]{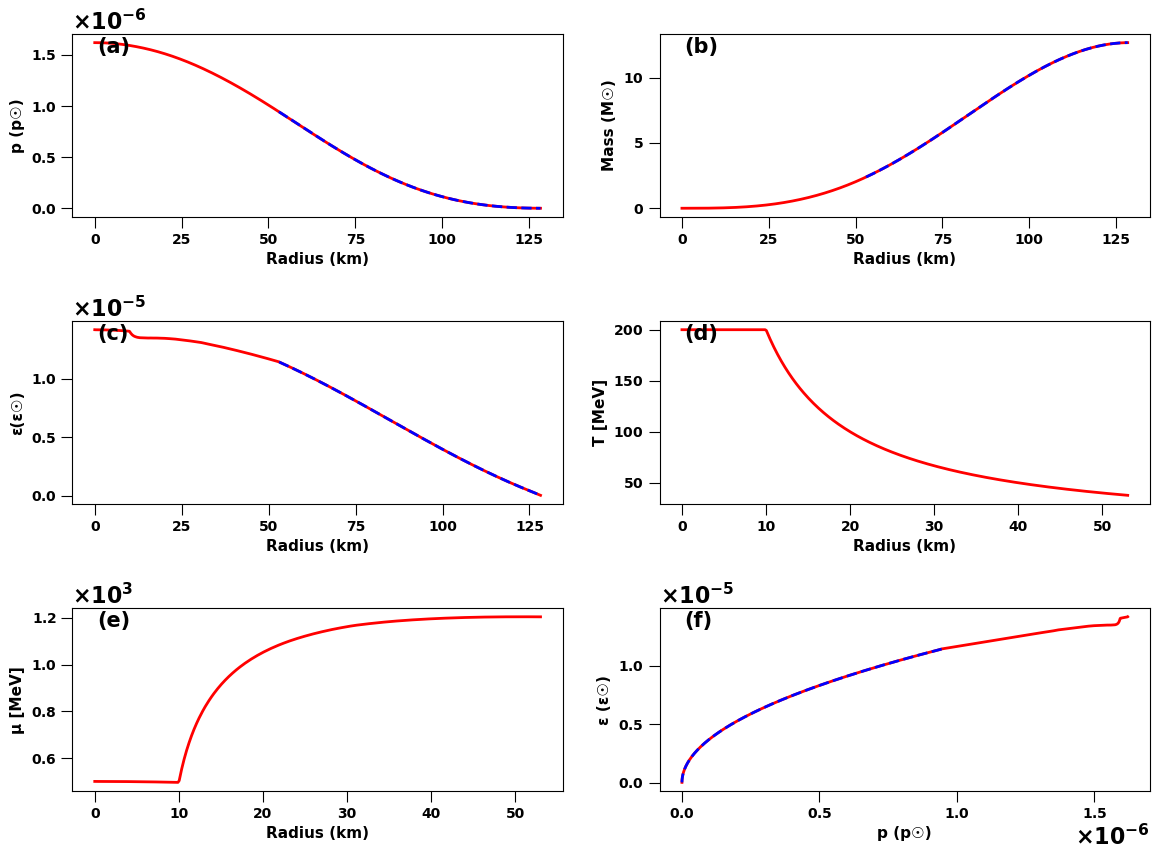}
    \caption{{Radial} 
 profiles of physical quantities for a complete QS model with $R_{\rm c} = 10\ \mathrm{km}$ and central parameters $(T_0, \mu_0) = (200\ \mathrm{MeV}, 500\ \mathrm{MeV})$. (\textbf{a}) Pressure, (\textbf{b}) mass accumulation, (\textbf{c}) energy density, (\textbf{d}) temperature, (\textbf{e}) chemical potential, and~(\textbf{f}) EoS. Blue dashed curves denote contributions from the external single-polytropic hadron matter EoS with $\gamma_{n}=0.5$, while red curves represent contributions from the internal quark matter.}
    \label{fig:Radial profiles}
\end{figure}

Figure~\ref{fig:Radial profiles} presents radial profiles of physical quantities for a single star of the complete QS model with $R_{\rm c} = 10\ \mathrm{km}$ and central parameters $(T_0, \mu_0) = (200\ \mathrm{MeV}, 500\ \mathrm{MeV})$. For~simplicity, but~enough to illustrate the idea, we select the case with an energy density jump factor of \( m = 1 \), which implies that the energy density is continuous at the phase transition.
Then the constructed QS has the quark core radius $R_t\approx53$ km, and~the total radius is 128 km, while the total mass is 12.7 $\text{M}_{\odot}$, which is indeed well within the range spanned by the five different EoS cases in Section~\ref{sec:2}. 

{Figure~\ref{fig:Radial profiles}a--c}, 
 radial profiles of pressure, cumulative mass, and~energy density. The~blue dashed curves denote contributions from the external hadron shell. Due to the imposed energy density jump factor \( m = 1 \) at the interface, both pressure and cumulative mass remain smooth across the phase transition boundary. Discontinuities in derivatives of the energy density occur at \( R_c \) and the phase transition~interface.  

{Figure~\ref{fig:Radial profiles}d,e}, the radial profiles of temperature \( T \) and chemical potential \( \mu \) within the star. The~temperature \( T \) strictly follows the distribution of (\ref{eq:T-profile}). We truncate our calculations of $T$ and $\mu$ at the interface $R_t$, since the EoS of the external hadronic phase is described by a simple single polytrope, where the dependencies of $T$ and $\mu$ are not explicitly~assigned. 

Within \( R_c \), the~temperature is uniform, but~the chemical potential \( \mu \) decreases along the constant \( T_0 \) line due to the pressure gradient term in the TOV equations. Beyond~\( R_c \), \( \mu \) initially increases but slightly decreases near the interface, intersecting the phase transition~line.

{Figure~\ref{fig:Radial profiles}f}, EoS of the star showing \( p(\epsilon) \) at different radii. The~blue segment corresponds to the external single-polytropic NS~EoS.

Under the condition of constant thermal conductivity, this section constructs the temperature- and chemical-potential-dependent EoS via a holographic two-flavor QCD model and~investigates the thermodynamics of QS through ``combined'' configurations with phase transition interfaces. Key findings highlight the unique EoS dependence on the central temperature \(T_0\) and the limiting behavior as \(R_c \to 0\). 

In the model of this section, to~circumvent the divergence of the central temperature, we introduced a central thermal core. However, we observed that its limiting case corresponds to a degenerate solution. Although~the model yields an approximate well-behaved conclusion, more precise results should be derived from a holographic framework in future~studies.

\section{Conclusions}\label{sec:6}

We extract the EoS for the 2-flavor quark--gluon plasma at a high temperature using a holographic QCD model. Since the real EoS is a complicated function of chemical potential and temperature, we select five different combinations of $\mu$ and $T$ near CEP for simplified analysis and employ them as the core of QS, which would be enough to enclose the real EoS and offer a quick estimation for the possible~range.

We first derive the M--R and the I--Love--Q--C relations with the ``simple'' 2-flavor quark EoS, which are analytical continuations of the fitting curves truncated at the phase transition points. Further to be more realistic and to make comparisons, we pay attention to ``combined'' QS: connecting the quark cores with polytropic hadron shells under different phase transitions and subsequently apply the ``combined'' cases to calculate the universal~relations. 

The results reveal that the mass distribution of both the ``simple'' and the ``combined'' QS are above 2 $\text{M}_{\odot}$, which enables them to be BH mimickers. But~the different wave forms and zero TLN of BH in the general relativity can help distinguish the QS from BH. The~mass and radius combinations we obtain from the 2-flavor EoS are much different from those from the 2+1-flavor models~\cite{Chen:2025caq}, providing a method to infer the fundamental compositions of the hot QS. Compared with the NS, the~I--Love--Q relations remain valid for the ``simple'' QS and the ``combined'' cases except for the $m=0.5$ model, extending the universal relations for both the continuous and the small discontinuous EoS. The~relations about compactness deviate more from each model than the I--Love--Q relations, providing us an effective method to constrain the phenomenological models with phase~transitions.

From Figures~\ref{iloveqc}--\ref{iloveqc05}, it can be observed that quark stars exhibit a broad mass coverage spanning from about $2$ to $17$ $\text{M}_{\odot}$, including the $2.5$--$5$ $\text{M}_{\odot}$ range, which may effectively interpret the so-called mass gap. Another very interesting outcome is that the~minimum mass  of QS (determined by the lowest pressure at the phase transition) approximates $2$ $\text{M}_{\odot}$. Below~this threshold mass, hot QS would undergo a phase transition to pure hot NS.  This critical mass closely aligns with the upper mass limit of NS, which should not be a pure coincidence. In fact, this shows consistency with the results in the literature~\cite{Hoyos:2016zke,Hoyos:2016cob,Annala:2017tqz,BitaghsirFadafan:2019ofb,Zhang:2022uin} that quark matter is not favored inside NS. If~one wants to obtain QS, the~masses should exceed $2$~$\text{M}_{\odot}$, in~addition to very high temperature.
Unlike QS, hot NS lack a radiation-reflecting interface, leading to rapid cooling processes and~result in normal cold NS under $2$ $\text{M}_{\odot}$. This thermal characteristic could explain the observational absence of hot NS possessing radii as large as those predicted in our~result.

An important new result of this work is presented in Figure~\ref{EoS}, though~not emphasized extensively,
where we numerically determine all the $T$, $\mu$ dependence for $p$ and $\epsilon$ of the quark phase in a holographic 2-flavor QCD model, presented as heatmaps spanning the entire parameter~space.

Finally, we also discuss the temperature- and chemical-potential-dependent EoS under constant thermal conductivity with a heat source. Starting from the Laplace equation and combined with the analysis of phase transition interfaces in QS, we reveal the thermodynamic properties of QS under this condition. It highlights the EoS dependence on the central temperature \(T_0\) and the limiting behavior as \(R_c \to 0\), providing insights for potential future~investigations.

Considering the model we refer to is based on lattice data with matched bare quark masses and a crossover temperature of $T_c(\mu_B=0)=205~\rm{MeV}$, whereas recent study~\cite{Borsanyi:2020fev}{ obtained a QCD phase transition line result based on lattice data with physical quark masses and a crossover temperature of $T_c(\mu_B=0)=158~\rm{MeV}$, we can use this model to fix parameters and potentially obtain a more realistic EoS, which will be left for future work.}


\vspace{6pt} 






\acknowledgments{The authors thank 
 Chian-Shu Chen, Alessandro Parisi, Niko Jokela, Yi-Zhong Fan, Qiyuan Pan, Shao-Feng Ge and an anonymous friend for very helpful suggestions and discussions.  K.Z. (Hong Zhang) is supported by a classified fund from Shanghai city.}




\end{document}